\documentclass[twocolumn,
superscriptaddress,amsmath, amssymb, amsfonts,preprintnumbers,aps,prd,
longbibliography,nofootinbib]{revtex4-2}

\usepackage{scalerel}
\usepackage{color}
\usepackage{amsmath,amsfonts,amssymb}%,calrsfs}
\usepackage{slashed}
\usepackage{latexsym,epsfig}
\usepackage{dsfont}
\usepackage{arydshln}
\usepackage{extarrows}
\usepackage{hyperref}
\usepackage{array}
\def\moth{\mathsurround=0pt}
\newdimen\zo \zo=0pt

\def\tick{\leaders\hrule height 0.5ex depth 0pt \hskip 0.5pt}
\def\upboxfill{$\moth \setbox\zo\hbox{\tick}%
  \hskip 3pt\hbox to 0pt{$\tick$\hss}\hrulefill \hbox to 7.5pt{$\tick$\hss}$}

\def\dtick{\leaders\hrule height .34pt depth 0.5ex \hskip 0.5pt}
\def\downboxfill{$\moth \setbox\zo\hbox{\dtick}%
  \hskip 2pt\hbox to 0pt{$\dtick$\hss}\hrulefill \hbox to 2pt{$\dtick$\hss}$}

\def\ov{\bar}

\def\bec{\begin{center}}
\def\ec{\end{center}}

\def\un{\underline}
\def\ov{\overline}

\def\cT{{\cal T}}

\def\nn{\nonumber}

\def\be{\begin{equation}}
\def\ee{\end{equation}}
\def\bea{\begin{eqnarray}}
\def\eea{\end{eqnarray}}
\def\ba{\begin{array}}
\def\ea{\end{array}}

\newcommand{\Exp}[1]{\operatorname{e}^{#1}}

\newcommand{\rmd}{{\mathrm{d}}}

\begin{document}

\title{On the inclusion of statistical matter in the non-relativistic limit of NS-NS supergravity}

\author{Eric Lescano} 
\email{eric.lescano@uwr.edu.pl}
\affiliation{University of Wroclaw, Faculty of Physics and Astronomy, Maksa Borna 9, 50-204 Wroclaw,
Poland}

% \date{\today}

\begin{abstract}
We combine techniques from kinetic theory and string dualities to couple statistical matter to a non-relativistic (NR) supergravity background, enabling the theory to be formulated in a T-duality invariant form. Similarly to the relativistic case, we explicitly demonstrate that, in the NR limit, the many-strings system necessitates a viscous fluid description for the statistical matter. This is achieved using the generalized energy-momentum tensor of a perfect fluid, which remains covariant under T-duality.
\end{abstract}

\maketitle

\section{Introduction}
T-duality \cite{T-duality1} is an exact symmetry \cite{Tduality,Tduality2} of string theory (ST) and its effect on the non-relativistic (NR) limit of ST 
\cite{NRintro1,NRintro2,NRintro3,NRintro4,NRintro5,NRintro6} is that a space-like longitudinal direction of the theory is mapped to a null one, underlying a discrete light-cone quantization \cite{Null1,Null2,Null3}. Remarkably, this leads to a drastic simplification of ST while its low energy limit is still given by a supergravity with a finite action principle \cite{Bergshoeff:2015uaa,Bergshoeff:2021bmc,tendimesional,Bidussi:2021ujm}. In the relativistic case, the energy-momentum tensor of a many-strings system is given by 
\bea
T^{ij} = \frac{1}{4\pi\alpha'}\int d\sigma \frac{d\tau}{dX^0} (\partial_{\tau}X^i \partial_{\tau}X^j - X'^i X'^j) \, , 
\eea
which can be effectively described  by a fluid with viscosity \cite{GasVene}, the latter appearing to compatibilize T-duality with the matter description, i.e.,
\bea
\tilde p = 0 \, , \quad \tilde \eta \neq 0 \, ,
\label{nop}
\eea
where $\tilde p$ is the pressure and $\tilde \eta$ is the share viscosity after a T-duality transformation. The physics behind this statistical model is the angular stone of string cosmology \cite{TVafa}.

In this work we will construct a T-duality invariant model for statistical strings under the NR limit of ST and we will show that the viscosity contribution is not required. The vacuum description of the NR supergravity is given by a transverse vielbein $\tau_{\mu}{}^{a}$, a longitudinal vielbein $e_{\mu}{}^{a}$, a two-form $b_{\mu \nu}$ and a scalar field $\phi$. All these fields and their dynamics can be rewritten in a Double Field Theory (DFT) \cite{DFT1,DFT2,DFT3,DFT4} formalism \footnote{For reviews see \cite{ReviewDFT1,ReviewDFT2,ReviewDFT3} and the second lecture in \cite{ReviewDFTE}.} where the fundamental fields obey a convergent c-expansion. Interestingly enough, in the supergravity case, all the bosonic fields contain divergences in their expansions and therefore the finitness of the action and the equations of motion can be easily recognized from the DFT prescription \cite{EyD}. Moreover, the DFT rewriting is a particular case of a larger family of non-Riemannian double geometries \cite{NRDFT1,NRDFT2,NRDFT3,NRDFT4,NRDFT5,NRDFT6,NRDFT7}. We can summarize these concepts in the following diagram,
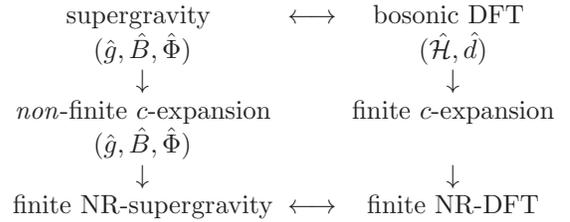
\begin{align}
    \begin{array}{ccccc}
      \textrm{supergravity }   &  \longleftrightarrow & \textrm{bosonic DFT }& &\\
      (\hat{g},\hat{B},\hat{\Phi})& & (\hat{\cal H},\hat{d}) & & \\
       \downarrow  & & \downarrow \\
       \text{\textit{non}-finite $c$-expansion} &   & \text{finite $c$-expansion}  \\
        (\hat{g},\hat{B},\hat{\Phi}) & &   \\
       \downarrow  & & \downarrow & & \\
       \textrm{finite NR-supergravity} & \longleftrightarrow & \text{finite NR-DFT}  & & 
    \end{array} \nonumber
\end{align}
where the $\leftrightarrow$-arrow means
\bea
& \rightarrow &\textrm{: DFT rewriting,} \nn \\
&\leftarrow &\textrm{: breaking the duality group.} \nn
\eea
The tension between the generalized frame formalism \cite{HK:frame} and generalized metric formalism \cite{Genmet} is also present when the generalized fields are expanded using a c-expansion \cite{Tension}. The former contains divergences in its expansion while the generalized metric is convergent. Therefore, both the action and the equations of motion converge, which indeed forces a very particular form for the generalized energy-momentum tensor. 

In this work we distinguish statistical matter from ordinary matter. The former is described by a distribution function which satisfies the Boltzmann equation while the latter is given by a matter Lagrangian. In the context of relativistic hydrodynamics, the equilibrium distribution function for a perfect fluid is given by \cite{Cercignani} 
\bea
f(x,p)\propto e^{-p_{\mu} \beta^{\mu}}
\eea
with $\beta^{\mu}= \frac{u^{\mu}}{T}$ and $T$ the temperature of the fluid. Then, the integration $\int_{x,p} g f(x,p) p^{\mu} p^{\nu}$ results in 
\bea
T^{\mu \nu}=(e+p) u^{\mu} u^{\nu} + p g^{\mu \nu} \, . 
\eea
While the inclusion of the perfect fluid in the NR supergravity is given by the Einstein equation
\bea
G_{\mu \nu} & = & k T_{\mu \nu} \, ,
\eea
with $\kappa= 8\pi G_{D}$, there is not a unique way to perform the c-expansion on the hydrodynamics variables. Therefore, in this work we propose to use DFT as a guide principle to construct the expansions and give the explicit form of energy-momentum tensor. This means that we will use the $O(D,D)$ symmetry to constrain the form of the expansion of the hydrodynamics variables. 

The inclusion of statistical matter in the DFT framework was given in \cite{EN1} and further study in \cite{EN2, EN3, ENY}. In these series of works, the generalized energy-momentum for a perfect fluid coupled to the double geometry was constructed from a kinetic theory perspective. This tensor is fully compatible with the generalized Einstein equation \cite{Parkcosmo1,Parkcosmo2}, and also is in correspondence \cite{Correspondence} with the dynamics of a massless generalized scalar field. 
The fluid description of the statistical matter constitutes, in general, a simple toy model to study the string hydrodynamics/thermodynamics and cosmological scenarios, among others. In those cases, the fluid might represent a string gas \cite{stringgas} or a more general many-strings system. The main goal of this work is to construct the explicit form of the energy-momentum tensor for these kind of scenarios, in such a way that the whole NR supergravity can be embedded in a T-duality invariant formalism. We will construct the energy-momentum tensor from the DFT framework in order to ensure that the final result is compatible with $O(D,D)$ symmetry. We will follow the same strategies implemented in \cite{NRmatter1,NRmatter2,NRmatter3,NRmatter4}, but starting from the double geometry and respecting the constraints of the latter. The main results of this work are:
\begin{itemize}

\item We construct the expansion of the generalized velocity $U_{M}$ in order to satisfy the usual DFT constraints $U_{M} {\cal H}^{M N} U_{N}=-1$, $U_{M} \eta^{M N} U_{N}=0$. We also verify that the generalized distribution function $f(X,{\cal P})$ \cite{ENY} remains finite after considering the NR limit. This result can be found in (\ref{sub1}) and (\ref{sub3}). 

\item We study the generalized correspondence \cite{EN3} when the fundamental fields are expanded under the NR expansion. This correspondence is compatible in the NR limit, giving rise to a matter Lagrangian $\rho^{(0)}(\Phi^{(0)})$. This result can be found in (\ref{sub2}).

\item We construct the generalized energy-momentum tensor for a perfect fluid, which leads to a fluid model in supergravity with dilaton source given by $\sigma = 2p$ in agreement with \cite{EN2}. This scenario leads to the viscosity inclusion given in \cite{ENY}. The explicit form of both the generalized energy-momentum tensor and its parametrization is the main topic of section (\ref{sub3}).
\end{itemize}

\section{Coupling matter in the non-relativistic formulation of NS-NS supergravity}
Our starting point is the standard NS-NS supergravity formulation with action principle given by
\begin{align}
\!\!\! S_{0} = \int \rmd^Dx\,\Exp{-2 \hat \Phi}\sqrt{-\hat g}\left[R(\hat e)+4\,(\partial \hat \Phi)^2 - \frac{1}{12}\,\hat H^2 \right], \nn
\end{align}
where the indices $\mu,\nu=0,\dotsc,D-1$ and $\hat H_{\mu \nu \rho}=3 \partial_{[\mu} \hat B_{\nu \rho]}$ and $\hat g$ is the determinant of the metric. The fundamental fields are expanded in powers of $c$ as
\bea
\hat e_{\mu}{}^{a} & = & c \ \tau_{\mu}{}^{a} \, , \quad \hat e_{\mu}{}^{a'} =  e_{\mu}{}^{a'} \, \\
\hat B_{\mu \nu} &=&  - c^2 \epsilon_{a b} \tau_{\mu}{}^{a} \tau_{\nu}{}^{b} + b_{\mu \nu} \, , \\
\hat{\Phi} &=& \ln(c) + \phi  \, ,
\eea
where $\hat a=(a,a')$ is the decomposition of the Lorentz index, $a=0,1$ are the transverse directions and $a'=2,\dots,D-1$ are the longitudinal ones. The new fundamental fields obey the following relations
\bea
  \tau_{\mu}{}^{a} e^{\mu}{}_{a'} & = & \tau^{\mu}{}_{a} e_{\mu}{}^{a'} = 0 \, , \qquad  e_{\mu}{}^{a'} e^{\mu}{}_{b'} = \delta^{a'}_{b'} \, , \\
  \tau_{\mu}{}^{a} \tau^{\mu b} & = & \eta^{a b} \, , \qquad \tau_{\mu}{}^{a} \tau^{\nu}{}_{a} + e_{\mu}{}^{a'} e^{\nu}{}_{a'} = \delta_{\mu}^{\nu} \, , \\
  h_{\mu \nu} & = & e_{\mu}{}^{a'} e_{\nu a'} \, , \qquad \qquad h^{\mu \nu}= e^{\mu a'} e^{\nu}{}_{a'} \, .
\eea

The limit $c\rightarrow \infty$ at the level of the supergravity Lagrangian is convergent, meaning that all the divergences coming from the Ricci scalar are cancelled by contributions coming from the curvature of the B-field. On the one hand, one can couple matter considering a variational principle. In that case, the Einstein equations can be schematically written as
\bea
\frac{2}{\sqrt{-g}}\frac{\delta S_{0}}{\delta h^{\mu\nu}} & = & T_{\mu \nu} \, , \ \
\frac{2}{\sqrt{-g}}\frac{\delta S_{0}}{\delta \tau^{\mu}{}_{a}} =  T_{\mu}{}^{a} \, , \\
\frac{2}{\sqrt{-g}}\frac{\delta S_{0}}{\delta b_{\mu\nu}} & = & J^{\mu \nu} \, ,  \ \ 
\frac{1}{\sqrt{-g}}\frac{\delta S_{0}}{\delta\phi} =  - \sigma \, ,
\eea
and the right hand side is given by the matter sources defined as
\bea
T_{\mu \nu} & = & \frac{-2}{\sqrt{-g}} \frac{\delta S_{\rm mat}}{\delta h^{\mu \nu}}\, , \ \  
T_{\mu}{}^{a} =  \frac{-2}{\sqrt{-g}} \frac{\delta S_{\rm mat}}{\delta \tau^{\mu}{}_{a}}\, , \ \ \\
J^{\mu\nu} & = & \frac{-2}{\sqrt{-g}}\frac{\delta S_{\rm mat}}{\delta b_{\mu\nu}}\, , 
\sigma =  \frac{1}{\sqrt{-g}}\frac{\delta S_{\rm mat}}{\delta\phi} \, ,
\eea
where $T_{\mu \nu}$ is the spacial energy-momentum tensor, $T_{\mu}{}^{a}$ is the energy current. On the other hand, one can directly couple statistical matter only considering the Einstein equation, 
\bea
G_{\mu \nu} & = & k T_{\mu \nu} \, ,
\eea
where $T_{\mu \nu}$ might be, for example, the energy-momentum for a perfect or imperfect fluid. Then, for a supergravity background which would be obtained after solving the strong constraint in the DFT framework upon considering the NR limit, we know that the lhs of the equation is finite due to the finiteness of the generalized fields. Therefore, the energy-momentum tensor requires the following expansion,
\bea
T_{\mu \nu} & = & T^{(0)}_{\mu \nu} + \mathcal{O}(\frac{1}{c^2}) \, .
\eea
In this work we will study this last scenario where the statistical matter is coupled directly considering the inclusion of the double perfect fluid in the framework of DFT and we will construct the compatible energy-momentum tensor using a NR expansion on the $O(D,D)$ multiplets.

\section{Construction of the model}

Recently \cite{EyD}, a formulation of NR DFT (compatible with the non-Riemannian formulation of DFT) has been constructed. The idea of this work is to include statistical matter to this model, in order to obtain statistical matter coupled to a NR supergravity upon parametrization. For simplicity we will restrict our study to the bosonic case, meaning that the duality group is $O(D,D)$. The invariant DFT metric is given by
\begin{align}
{\eta}_{{M N}} = \left(\begin{matrix}0&\delta_\mu^\nu\\ 
\delta^\mu_\nu&0 \end{matrix}\right)\,, \label{etaintro}
\end{align}
with $M,N,\dots=0,\dots,2D-1$ in the fundamental of the duality group, while the generalized metric is expanded as
\bea
\hat {\cal H}_{M N} = {\cal H}^{(0)}_{M N} + \frac{1}{c^2}{\cal H}^{(-2)}_{M N} \, ,
\eea
where we use the suffix (i) to indicate the order of a given term. Moreover, in order to match with the NR limit of supergravity, the parametrization of the previous metric is given by
\begin{widetext}
\begin{align}
    {\cal H}^{(0)}_{M N} &=  
\left(\begin{matrix} h^{\mu \nu} &  \epsilon_{a b} \tau_{\nu}{}^{b} \tau^{\mu a}  - b_{\rho \nu} h^{\rho \mu} \\ 
\epsilon_{a b} \tau_{\mu}{}^{b} \tau^{\nu a}  - b_{\rho \mu} h^{\rho \nu} &  h_{\mu \nu} + b_{\rho \mu} h^{\rho \sigma} b_{\sigma \nu} - 2 \epsilon_{c d}  \tau_{(\mu|}{}^{d} b_{\sigma |\nu)} \tau^{\sigma c} \, \end{matrix} \right) \quad
    {\cal H}^{(-2)}_{M N} &=  
\left(\begin{matrix} \tau^{\mu a} \tau^{\nu}{}_{a} &  -b_{\rho \nu} \tau^{\rho c} \tau^{\mu}{}_{c} \\ 
-b_{\rho \mu} \tau^{\rho c} \tau^{\nu}{}_{c} &  b_{\rho \mu} \tau^{\rho c} b_{\sigma \nu} \tau^{\sigma}{}_{c} \end{matrix}\right)  \, .
\end{align}
\end{widetext}
As we can easily observe, the main difference between NR DFT and NR supergravity is that in the former, the dynamical metric is finite while in the latter the expansion of the metric (and the expansion of the B-field) contains divergences.

As in the relativistic case, a crucial consistency equation of the DFT formulation is given by the strong constraint 
\begin{align}
\partial_{M} (\partial^{M} \star) = 0 \, \quad (\partial_{M} \star) (\partial^{M} \star) = 0 \,,
\label{SC}
\end{align}
where $\partial_{M}=(\partial_{\mu},\tilde{\partial}^\mu)$ refers to derivatives with respect to the ordinary/dual coordinates, respectively, and $\star$ represents any combination of generalized fields and/or gauge parameters. Thanks to the strong constraint the closure of the generalized diffeomorphism transformations infinitesimally  by $\xi^{M}$  through the generalized Lie derivative
\bea
{\cal L}_\xi V_M = \xi^{N} \partial_N V_M + (\partial_M \xi^N - \partial^N \xi_{M}) V_N  \, ,
\eea 
is guaranteed. The DFT Jacobiator is not trivial and therefore the algebraic structure of DFT is given by an $L_{\infty}$-algebra with a non-trivial $l_3$ product \cite{Linf1}-\cite{Linf4}, which measures the failure of the Jacobi identity in the double geometry. The usual solution to (\ref{SC}) in order to recover the supergravity framework is $\tilde{\partial}^\mu=0$\,. In this formulation the Lorentz transformations act trivially, but we refer to \cite{EyD} for details on the parametrization of the double Lorentz parameter. 

On the other hand, it is worth observing that ${\cal H}^{(0)}_{M N}$ is an element of the duality group, i.e.,
\bea
{\cal H}^{(0)}_{M P} \eta^{P Q} {\cal H}^{(0)}_{Q N} & = & \eta_{M N} \, .
\eea
Therefore, we can construct the leading order of the generalized projectors in the following way,
\bea
P^{(0)}_{MN} & = & \frac{1}{2}\left(\eta_{MN} - {\cal H}^{(0)}_{MN}\right) \, ,  \\
\ov{P}^{(0)}_{MN} & = & \frac{1}{2}\left(\eta_{MN} + {\cal H}^{(0)}_{MN}\right)\ ,
\eea
and they satisfy the following properties 
\bea
&{\overline{P}}^{(0)}_{{M Q}} {\overline{P}}^{(0) Q}{}_{ N}={\overline{P}}^{(0)}_{{M N}}\, , &\quad {P}^{(0)}_{{M Q}} {P}^{(0)Q}{}_{ N}={P}^{(0)}_{{M N}}, \nn\\
&{P}^{(0)}_{{M  Q}}{\overline{P}}^{(0)Q}{}_{ N} = {\overline{P}}^{(0)}_{ {M Q}}  {P}^{(0) Q}{}_{ N} = 0\, ,  &\quad {\overline{P}}^{(0)}_{{MN}} + {P}^{(0)}_{{M N}} = \eta_{{M N}}\,. \nn
\eea

The generalized dilaton has a finite $c$-expansion, as the expansion of the dilaton $\hat{\Phi}$ and the metric determinant $\hat g$ cancel,
\bea
e^{-2 \hat{d}} = e^{-2\hat{\Phi}} \sqrt{-\hat g} = e^{-2 \phi} \sqrt{f(\tau,h)} = e^{-2 d},
\label{dilaton}
\eea 
where $f(\tau,h) = - \frac{\hat{g}}{c^4}$.

Since both $\hat {\cal H}_{M N}$ and $\hat d$ do not contain divergences, all the leading contributions to the generalized curvatures is given by ${\cal H}^{(0)}_{M N}$ and $d$, i.e., 
\bea
{\cal R} & = & \frac18 {\cal H}^{(0) M  N} \partial_{ M}{\cal H}^{(0) K  L}\partial_{ N}{\cal H}^{(0)}_{ K  L} + 4 {\cal H}^{(0) M  N} \partial_{ M}d \partial_{ N}  d\nn \\ && - \frac12 {\cal H}^{(0) M  N}\partial_{ N}{\cal H}^{(0) K  L}\partial_{ L}{\cal H}^{(0)}_{ M  K} \nn \\ &&  - 2 (\partial_{ M}{\cal H}^{ (0)M  N}) \partial_{ N} d + \mathcal{O}\left(\frac{1}{c}\right) \, , \label{cLagrangian}
\eea
and
\bea
{\cal R}_{M N} = P^{(0)}_{M}{}^{P} {\cal K}_{P Q} {\overline P}^{(0)Q}{}_{N} +  {\overline P}^{(0)}_{M}{}^{P} {\cal K}_{P Q} P^{(0)Q}{}_{N} \, ,\label{riccitensor}
\eea
with
\bea
{\cal K}_{M N} & = & \frac{1}{8} \partial_{M} {\cal H}^{(0)K L} \partial_{N} {\cal H}^{(0)}_{K L} \nn \\&& - \frac14 \left(\partial_{L} - 2 \partial_{L} d\right)\left({\cal H}^{(0)L K} \partial_{K}{\cal H}^{(0)}_{M N}\right) \nn \\ && + 2 \partial_{M}\partial_{N} d  - \frac12 \partial_{(M} {\cal H}^{(0)K L} \partial_{L} {\cal H}^{(0)}_{N) K} \nn \\ && + \frac12 \left(\partial_{L} - 2 \partial_{L} d\right) \left({\cal H}^{(0)K L} \partial_{(M} {\cal H}^{(0)}_{N) K} \right. \nn \\ && \left. + {\cal H}^{(0)K}{}_{(M} \partial_{K} {\cal H}^{(0)L}{}_{N)}\right) \, . \nn 
\eea

The generalized Einstein equation is given by \cite{Parkcosmo1,Parkcosmo2},
\begin{align}
\mathcal{G}_{MN}= k \mathcal{T}_{MN}\label{einstein} \,,
\end{align}
with the generalized Einstein tensor defined as
\begin{align}
\mathcal{G}_{MN}=\mathcal{R}_{MN}-\frac12\,\mathcal{R}\,\hat {\cal H}_{MN} \, .
\end{align}
Considering that the generalized energy-momentum tensor can be factorized as
\begin{align}
\mathcal{T}_{MN}=\hat{\mathcal{T}}_{MN}-\frac12 \,\cT\,\hat{\cal H}_{MN}\,,\label{tmntotal}
\end{align}
where both ${\cal R}_{MN}$ and $\hat{\cT}_{MN}$ contain only mixed components with respect to the full DFT projectors, then (\ref{einstein}) reads
\begin{align}
\mathcal{R}_{MN}&= \kappa\hat{\mathcal{T}}_{MN}\,,\label{rmneq}
\\
\mathcal{R}&= \kappa \mathcal{T}\label{req}\,.
\end{align}

The explicit form of the generalized energy-momentum tensor for a perfect fluid configuration is given by
\cite{EN2} \footnote{This tensor is a particular case of the recent proposal of \cite{ENY}, where the authors included non-perfect terms as well. Also, it is the commutative version of the one in \cite{EyT}.},
\bea
 {\mathcal T}_{MN}= (e+p) \left( \hat U_{\ov { M}}\, \hat U_{\un { N}}+ \hat U_{\ov { N}}\, \hat U_{\un { M}}\right)+ p \,\hat{\cal H}_{ MN}\,,
\label{proposaltmn}
\eea
where $\hat U_{M}$ is a generalized velocity which satisfies
\bea
\label{constU}
\hat U_{M} \hat {\cal H}^{M N} \hat U_{N} & = & - 1 \\
\hat U_{M} \eta^{M N} \hat U_{N} & = & 0 \, .
\label{constU2}
\eea
Since T-duality relates perfect fluid with imperfect fluids \cite{GasVene}, the supergravity reduction of the previous generalized energy-momentum tensor induces viscosity terms \cite{ENY}. 

The constraints (\ref{constU})-(\ref{constU2}) ensure that the generalized velocity describes the ordinary D-velocity of the fluid hydrodynamics. From the previous relations it is natural to follow a similar expansion to the one in \cite{NRmatter3}, but in the double geometry,
\bea
\hat U_{M} = U^{(0)}_{M} + \mathcal{O}(\frac{1}{c}) \, .
\label{Genvel}
\eea
The previous expansion requires a particular form for the parametrization of $U^{(0)}$ since the generalized velocity needs to satisfy the constraints (\ref{constU}), which can be written as
\bea
\label{c1}
U^{(0)}_{M} {\cal H}^{(0) M N} U^{(0)}_{N} \, 
 & = & 0 \, , \\
U^{(0)}_{M} \eta^{M N} U^{(0)}_{N} & = & 0 \, .
 \label{c2}
 \eea

Besides (\ref{c1})-(\ref{c2}), we need to ensure that the generalized distribution function related to double phase space/kinetic theory formalism is also finite. We will explore this point in the next subsection. Finally, at this point we have learned information about the hydrodynamics quantities involved in the generalized energy-momentum tensor. Since the generalized velocity contains $c^2$ contributions and the projectors are at most order $1$, then the generalized energy density and pressure admit an expansion of the form,
\bea
\label{eyp}
e & = & e^{(0)} + \mathcal{O}(\frac{1}{c}) \, , \quad p = p^{(0)} + \mathcal{O}(\frac{1}{c}) \, .
\eea
These expansions are not in agreement with the generalized correspondence between a perfect fluid and a scalar field, so we will briefly discuss it in the subsection (\ref{sub2}). 

\subsection{Statistical point of view}
\label{sub1}
The statistical description of the perfect fluid was originally described in \cite{EN1}. In this section we inspect the conditions to obtain a finite distribution function for the perfect fluid when the fields are expanded using the NR ansatz. 

The double phase space coordinates are given by  
\bea
\Big\{X^{M},{\cal P}^{M} \Big\} \, ,
\eea
where ${\cal P}^{M}$ is an $O(D,D)$ vector. The infinitesimal generalized diffeomorphisms acting on a double phase space vector $V^Q=V^{Q}(X,{\cal P})$ are defined as
\bea
\delta_{\xi} V^{Q} = {\cal L}_{\xi} V^{Q} + {\cal P}^{N} \frac{\partial \xi^{M}}{\partial X^{N}} \frac{\partial V^Q}{\partial {\cal P}^{M}} \, ,
\label{transintro}
\eea
where ${\cal L}_{\xi}$ is the generalized Lie derivative and $\xi_M=\xi_M(X)$ is an infinitesimal parameter. The section condition is given by
\bea
(\frac{\partial}{\partial {\cal P}^{M}} \star) (\frac{\partial}{\partial {\cal P}_{M}} \star) & = & \frac{\partial}{\partial {\cal P}^{M}} (\frac{\partial}{\partial {\cal P}_{M}} \star) = 0 \,  \\
(\frac{\partial}{\partial X^{M}} \star) (\frac{\partial}{\partial {\cal P}_{M}} \star) & = & \frac{\partial}{\partial X^{M}} (\frac{\partial}{\partial {\cal P}_{M}} \star) = 0 \, . \,   
\eea
We solved the strong constraint of the double phase space with the solution
\bea
\frac{\partial}{\partial {\cal P}_{\mu}} = 0 \, . 
\eea
and ${\cal P}_{M}=(p^{\mu}, 0)$. The closure of (\ref{transintro}) is given by the C-bracket, and the covariant derivative of the double phase space is given by the generalized Liouville operator $D_{M}$ defined as 
\bea
D_{M} = \nabla_{M} - \Gamma_{M N}{}^{Q} {\cal P}^{N} \frac{\partial}{\partial {\cal P}^{Q}} \, .
\eea
The generalized Boltzmann equation considering a generalized distribution function $F=F\big( X,{\cal P} \big)$ is given by \cite{EN1}, 
\bea
{\cal P}^{M} {\cal D}_{M} F = {\cal C}[F] \, ,
\label{BoltzmannDFTintro}
\eea
where  ${\cal C}[F]$ is a generalized collision term and the operator ${\cal D}_M$ contains a generalized dilaton dependence,
\bea
{\cal D}_{M} = D_{M} - {\cal U}_{M} \, ,
\eea
with ${\cal U}_{M}= {\cal U}_{M}(X) = 2\,\partial_M \hat d.$  

The generalized equilibrium distribution function  is given by \cite{ENY},
\begin{align}
F_{\rm eq}(X,{\cal P})= \Exp{2\hat d - {\cal P}^{M} \hat {\cal H}_{M N} \beta^{N}}
\label{MBDFT}
\end{align}
where $\beta_{N} = \hat U_{N}/T$. Considering the NR expansion on the generalized dilaton and the generalized metric, and a generic expansion for the velocity we find
\begin{align}
F_{\rm eq}(X,{\cal P})= \Exp{2 d - \frac{1}{T}{\cal P}^{M}({\cal H}^{(0)}_{M N} + {\cal H}^{(-2)}_{M N}) U^{N}} \nn \, .  
\end{align}
On the other hand, the expansion of the generalized momentum can be induced from the expansion of the generalized metric,
\bea
{\cal P}_{M} = {\cal P}^{(0)}_{M} + \frac{1}{c^2}{\cal P}^{(-2)}_{M} \, ,
\eea
and the expansion of the generalized velocity is given by (\ref{Genvel}). Therefore, we find that the generalized distribution function is finite when $c\rightarrow\infty$. In this work we are not considering an expansion for the generalized temperature, but we can easily see that the temperature may include divergences that do not affect the behaviour of the generalized distribution function. At this point we have found the sufficient condition to construct a finite kinetic theory, and we will continue our discussion about the distribution function after parametrizing the fields, in section \ref{sub3}. In the next subsection we will study the generalized correspondence between the perfect fluid and a generalized scalar field in the double geometry for the NR case.

\subsection{Generalized correspondence with the scalar field}
\label{sub2}
The scalar field-perfect fluid correspondence \cite{Correspondence} is a formal correspondence between the energy-momentum tensor of a massless scalar field and the energy-momentum tensor of a perfect fluid. The latter can be constructed from the former considering a map between the scalar field and the hydrodynamics quantities. From this relation it is possible to write a matter Lagrangian in terms of the fluid variables. Particularly, ${\cal L}_{\rm m}= p$, where $p$ is a function of the scalar field. 

The generalization of the fluid velocity and scalar field derivative correspondence at the double geometry level is given by
\bea
\hat U^{(0)}_{ M} = \frac{\partial_{ M}  \Phi^{(0)}}{\sqrt{|\hat{\cal H}^{(0)PQ} \partial_{P}\Phi^{(0)} \partial_{Q} \Phi^{(0)}}|} \, ,\label{velocitytoscalarfield}
\eea
while the generalized pressure and energy density are given by
\bea
\label{pcorrespondence}
p^{(0)} & = & - \frac12 \hat{\cal H}^{(0)PQ} \partial_{P}\Phi^{(0)} \partial_{Q} \Phi^{(0)} - V(\Phi^{(0)}) \, , \\ 
e^{(0)} + p^{(0)} & = & |\hat{\cal H}^{(0)PQ} \partial_{P}\Phi^{(0)} \partial_{Q} \Phi^{(0)}| \, .
\eea

These relations turn 
\bea
 {\cal T}_{ M  N} &= & - 4\, \overline{P}^{(0)}_{ K( M}\,P^{(0)}_{ N) L}\,\partial^{ K}  \Phi^{(0)} \partial^{ L} \Phi^{(0)} \nn \\ && - \frac12 \hat{\cal H}^{(0)}_{M  N}  \hat{\cal H}^{(0) R  Q} \partial_{ R} \Phi^{(0)} \partial_{ Q} \Phi^{(0)} \nn \\&& - \hat{\cal H}^{(0)}_{MN}\,V(\Phi^{(0)})\, .
\label{ScalarTMN}
\eea
into (\ref{proposaltmn}) in the NR limit. The previous identifications between the leading order contribution for the pressure, density energy and velocity prove that the dynamics of the generalized scalar field under the NR limit is capable of describir a NR fluid in the double geometry. Therefore, the matter Lagrangian is given by (\ref{pcorrespondence}), in agreement with the relativistic case \cite{EN3}.
 
\section{Breaking the duality group: the supergravity model}
\label{sub3}

The parametrization of the generalized velocity is given by
\bea
U^{(0)}_{\mu} = - b_{\mu \rho} u^{\rho} , \quad U^{(0) \mu}= u^{\mu} \, ,
\label{Genvelparam2}
\eea
where $u^{\mu}$ is the ordinary D-velocity. On the other hand, the decomposition of (\ref{tmntotal}) gives
\begin{align}
\hat{\cT}_{MN} & = (e+p)\left( \hat U_{\ov { M}}\, \hat U_{\un { N}}+ \hat U_{\ov { N}}\, \hat U_{\un { M}}\right) \,, 
\label{hatcT-p}\\
\mathcal{T} & = -2\,p \, .
\label{pressure}
\end{align}

After considering the NR limit, the generalized Einstein equation is
\bea
\label{genEinstein}
\mathcal{R}_{MN} & = & \kappa  (e^{(0)}+p^{(0)})\left( U^{(0)}_{\ov { M}}\, U^{(0)}_{\un { N}}+ U^{(0)}_{\ov { N}}\, U^{(0)}_{\un { M}}\right) \, ,
\eea
where the indices are projected with $P^{(0)}_{M N}$ and $\overline P^{(0)}_{M N}$. From the last equation we learn that the source for the dilaton is $\sigma=2p$ like in the relativistic case. When exploring the independent components of the rhs of (\ref{genEinstein}) considering $b_{\mu \nu}=0$ we obtain
\bea
\label{model}
\hat {\cal T}^{(0)}_{\mu \nu} & = & -\frac{1}{2} (e^{(0)}+p^{(0)}) (h_{\mu \rho} h_{\nu \sigma} u^{\rho} u^{\sigma}) \, , \\
\hat {\cal T}^{(0)}_{\mu}{}^{\nu} & = & \hat {\cal T}^{(0) \mu \nu} = 0 \, . 
\eea
As studied in \cite{EN2}, the b-field contributions are relevant beyond the dilaton gravity scenario, especially towards the construction of a NR cosmological model. The b-contributions to the energy-momentum tensor are
\bea
T_{\mu \nu}|_{b} & = & -\frac{1}{2} \kappa (e^{(0)}+p^{(0)})(- 2 b_{\epsilon \mu} h^{\epsilon \xi} b_{\lambda \nu} h^{\lambda \chi} b_{\xi \rho} b_{\chi \sigma} u^{\rho} u^{\sigma} \nn \\
&& + 4 \epsilon_{cd} \tau_{(\nu|}{}^{d} b_{\lambda \sigma} \tau^{\lambda c} b_{\epsilon |\mu)} h^{\epsilon \xi} b_{\xi \rho} u^{\rho} u^{\sigma} \nn \\ 
&& +2 \epsilon_{cd} \tau_{\rho}{}^{d} b_{\xi(\nu|} \tau^{\xi c} b_{\epsilon |\mu)} h^{\epsilon \sigma} b_{\sigma \lambda} u^{\lambda} u^{\rho} \nn \\ 
&& -\frac12 \epsilon_{ab} \tau_{\rho}^{b} b_{\xi \mu} \tau^{\xi a} \epsilon_{c d} \tau_{\sigma}{}^{d} b_{\lambda \nu} \tau^{\lambda c} u^{\rho} u^{\sigma} \nn \\ 
&& -2 \epsilon_{cd} \epsilon_{ab} \tau_{\rho}^{d} \tau^{\xi c} \tau_{(\nu|}{}^{b} \tau^{\lambda b} b_{\xi |\mu)} b_{\lambda \sigma} u^{\rho} u^{\sigma} \nn \\
&& -2 \epsilon_{ab} \tau_{\mu}{}^{b} \tau^{\rho a} \epsilon_{cd} \tau_{\nu}{}^{d} \tau^{\sigma c} b_{\lambda \rho} b_{\delta \sigma} u^{\delta} u^{\lambda} \nn \\
&& + \frac12 b_{\mu \rho} u^{\rho} b_{\nu \sigma} u^{\sigma} -2 h_{(\mu \rho} h^{\epsilon \sigma} b_{\sigma \xi} b_{\epsilon \nu)} u^{\xi} u^{\rho} \nn \\
&& + h_{(\mu| \rho} \epsilon_{cd} \tau_{|\nu)}{}^{d} \tau^{\lambda c} b_{\lambda \sigma} u^{\rho} u^{\sigma} \nn \\ && + h_{(\mu| \rho} \epsilon_{cd} \tau_{\sigma}{}^{d} \tau^{\lambda c} b_{\lambda |\nu)} u^{\rho} u^{\sigma} \nn \\ && + h_{(\mu| \rho} \epsilon_{a b} \tau_{|\nu)}{}^{b} \tau^{\sigma a} b_{\sigma \xi} u^{\rho} u^{\xi}) \, . 
\eea
When we compare this model with the ordinary energy-momentum tensor for imperfect fluids \cite{ENY}
\bea
T_{\mu \nu} = (e+p) \hat u_{\mu} \hat u_{\nu} + p \hat g_{\mu \nu} + \tilde \eta \Pi_{\mu \nu} \, ,
\eea
where $\Pi_{\mu \nu}$ is a general viscous stress tensor, we infer that the T-duality invariant model is constructed considering,
\bea
\hat u_{\mu} & = & h_{\mu \nu} u^{\nu} + \mathcal{O}(\frac{1}{c}) \, ,  \\ 
e & = & - \frac12 e^{(0)} + \mathcal{O}(\frac{1}{c}) \, , \\
p & = & - \frac12 p^{(0)} + \mathcal{O}(\frac{1}{c}) \, , \\
\Pi_{\mu \nu} & = & - \frac{1}{\tilde \eta}p^{(0)} h_{\mu \nu} + \mathcal{O}(\frac{1}{c})\, ,
\eea
with $p^{(0)} - \frac{\sigma}{2} = \tilde p=0$. The whole construction is in agreement with \cite{GasVene}, indicating that the generalized energy momentum-tensor requires imperfect contributions to embed it it the generalized fluid form, as briefly presented in (\ref{nop}) and further studied in \cite{ENY}. 

This shows that the rewriting of a NR string cosmology in terms of the DFT framework will need the use of imperfect terms in the generalized energy momentum tensor to cancel this contribution (and the ones coming from the b-field). We leave this complete analysis for future work.  

\section{Conclusions}
The generalized metric formalism of DFT is a useful framework to explore the NR limit of NS-NS supergravity due to its simplicity and the possibility to take the limit $c \rightarrow \infty$ before breaking the duality group. In this work we have extended this formalism and we have included statistical matter following the prescription given in \cite{EN1}-\cite{EN3}. Once more, T-duality proves to be a powerful symmetry to construct effective models related to string theory. The finiteness of the generalized Ricci tensor and scalar forces a particular expansion for the generalized energy-momentum tensor. The generalized velocity, in the NR limit, solves the typical DFT constraints ($\hat U_{M} \hat H^{M N} \hat U_{N}=-1$, $\hat U_{M} \eta^{M N} \hat U_{N}=0$) and also requires to be compatible with a finite generalized distribution function. All these constraints are solved considering the expansion (\ref{Genvel}) and the parametrization (\ref{Genvelparam2}). Using this finite expansion, the generalized energy density and pressure have to be expanded in a convergent form according to (\ref{eyp}). In this work we proved that the inclusion of statistical matter in terms of a double perfect fluid does not introduce divergences in the double formalism. We then summarize our findings in the following diagram,
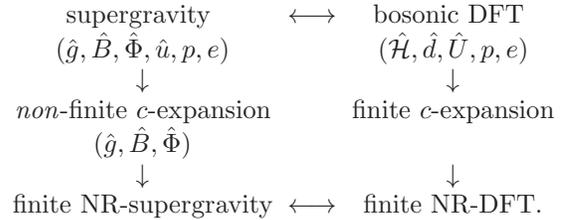
\begin{align}
    \begin{array}{ccccc}
      \textrm{supergravity }   &  \longleftrightarrow & \textrm{bosonic DFT }& &\\
      (\hat{g},\hat{B},\hat{\Phi},\hat u,p,e)& & (\hat{\cal H},\hat{d},\hat U,p,e) & & \\
       \downarrow  & & \downarrow \\
       \text{\textit{non}-finite $c$-expansion} &   & \text{finite $c$-expansion}  \\
        (\hat{g},\hat{B},\hat{\Phi}) & &   \\
       \downarrow  & & \downarrow & & \\
       \textrm{finite NR-supergravity} & \longleftrightarrow & \text{finite NR-DFT.}  & & 
    \end{array} \nonumber
\end{align}

We also prove that the generalized perfect fluid-scalar field correspondence is valid when we consider the NR expansion on the fundamental fields, which opens the possibility to create a NR DFT cosmology coupling a generalized scalar field which mimics the dynamics of the fluid in this limit. Finally, the main result of this work is the construction of the model (\ref{model}). Similarly to what happens in the relativistic case, the energy-momentum tensor of the perfect fluid requires imperfect terms in order to be embed in terms of the generalized perfect fluid. The results of this work pave the way to the study of NR cosmological backgrounds in ST following the prescription \cite{EN2} but coupling the matter contributions and using the generalized correspondence as established in this work.

\subsection*{Acknowledgements}
We thank  D. Marques for his comments on the first version of the draft, and N. Miron-Granese for enlightening discussions. We are in debt with our anonymous referee for point out the important re-scaling of the c parameter in the generalized Einstein equation, which is a crucial difference in equation (15) between the current version and the V1 of this work. The author is supported by the SONATA BIS grant 2021/42/E/ST2/00304 from the National Science Centre (NCN), Poland.

\end{document}